# MiBoard: Metacognitive Training Through Gaming in iSTART (April 2009)

Justin F. Brunelle, Irwin B. Levinstein, and Chutima Boonthum, *Member, IEEE*

*Abstract*—MiBoard (Multiplayer Interactive Board Game) is an online, turn-based board game, which is a supplement of the iSTART (Interactive Strategy Training for Active Reading and Thinking) application.  MiBoard is developed to test the hypothesis that integrating game characteristics (point rewards, game-like interaction, and peer feedback) into the iSTART trainer will significantly improve its effectiveness on students' learning.  It was shown by M. Rowe that a physical board game did in fact enhance students' performance.  MiBoard is a computer-based version of Rowe's board game that eliminates constraints on locality while retaining the crucial practice components that were the game's objective. MiBoard gives incentives for participation and provides a more enjoyable and social practice environment compared to the online individual practice component of the original trainer.

## I. MINTRODUCTION

iBOARD (MULTIPLAYER INTERACTIVE BOARD GAME) IS the computerized game version of Mike Rowe's physical iSTART board game. iSTART is a web-based tutor for high school students to improve their reading and thinking skills that includes an extended practice component. Currently, students are guided through extended practice modules which provide repetitive practice using the iSTART strategies (see *iSTART*) to create self-explanations. Students are given a text in which they are to create self-explanations for each of several targeted sentences. Multiple texts are given to the students on which to practice.

Research with iSTART has indicated the need for students to have extended practice with reading strategies.  This is because the effects of the initial iSTART training tend to taper over time and less skilled readers appear to need more training to achieve higher levels of comprehension (see *Evidence iSTART Works*).  Therefore, students need additional, extended practice after the initial training. This training takes time and practice. Unfortunately, research has also indicated that iSTART, while relatively engaging for most students initially, can become tedious for some students after a period of time. Perhaps this is because its layout is somewhat static, or possibly because the interactions (during extended practice) remain generally the same. A decline in engagement over time may also result from the lack of explicit incentive for the students to achieve mastery of the reading strategies. Rowe showed that a game can be used to help alleviate the tedium (see *iSTART: the Board Game*). MiBoard addresses all of the above concerns, including the lack of engagement.

MiBoard is an extension of iSTART that allows students to practice the skills targeted by iSTART in a more engaging and stimulating environment, while collaborating with their peers in a more social and structured educational forum. MiBoard is a 3- or 4-player turn-taking board game that gives players practice in making and analyzing self-explanations of sentences that occur in the context of longer texts.

The current extended practice emphasizes repetitively creating self-explanations with any of the strategies, while MiBoard emphasizes analytically creating self-explanations using a single, targeted strategy at a time and identifying the use of various strategies in peer self-explanations.

Experiments will be conducted using iSTART as it currently exists and iSTART using MiBoard in place of extended practice. MiBoard will be compared and contrasted to the current iSTART extended practice with respect to the effectiveness of MiBoard as a learning tool, as well as determine whether or not the students remain more engaged and have more fun while practicing using MiBoard as opposed to the existing extended practice.

## II. iSTART

iSTART (Interactive Strategy Training for Active Reading and Thinking) is a web-based tutoring system for high-school students that aims at teaching the users how to better understand science texts and textbooks [1]. Though the primary domain of the project is science, the skills acquired through the iSTART program can be applied to other areas, such as literature or history. Science texts are targeted because of the inherently complex nature of such compositions; scientists often use difficult concepts, complex sentences, and references to remote sentences when composing a text. In addition these compositions often contain technical jargon that make the text foreign to every-day experience and difficult for young adults to comprehend.

iSTART, developed with funding from the National Science Foundation and the Institute of Education Science, aims to provide instruction in reading strategies to support the process of self-explanation or explaining a poorly-

Manuscript received February 27, 2009. This work is supported by the National Science Foundation under Grant IIS-0725682. The original iSTART project was supported by the National Science Foundation Grant REC0241144 and its continuation was supported by the Institute of Education Science Grant R305G040046.

J. F. Brunelle, is a first-year graduate student in the Department of Computer Science, Old Dominion University, Norfolk, VA 23529 USA (e-mail: jbrunelle@cs.odu.edu).
I. B. Levinstein is with the Department of Computer Science, Old Dominion University, Norfolk, VA 23529 USA (e-mail: ibl@cs.odu.edu).
C. Boonthum is with the Department of Computer Science, Hampton University, Hampton, VA 23668 USA (e-mail: chutima.boonthum@hamptonu.edu).

comprehended sentence to oneself. Students who self-explain text are more successful at solving problems, more likely to generate inferences, construct more coherent mental models, and develop a deeper understanding of the concepts covered in the text [2], [3]. iSTART exposes the user to several strategies to be used during reading to enable the reader to better comprehend and retain the information being read. These strategies are explained in *iSTART Reading Strategies*. In iSTART the users type in their self-explanations about the texts being read. The tutoring system uses animated pedagogical agents to train the user in the use of those self-explanation strategies and other active reading strategies to explain (and therefore, comprehend) science texts.

Reading strategy instruction occurs in three stages with each stage requiring increased interaction on the part of the learner. During the Introduction Module of iSTART, the trainee is interactively engaged by a trio of animated characters that interact with each other by providing information, posing questions, and providing explanations of self-explanation and the reading strategies mentioned in the following sub-section, *iSTART Reading Strategies*. The three characters (an instructor and two students) speak using a text-to-speech synthesizer and a repertoire of gestures.

In the second phase, called the Demonstration Module, two agents demonstrate the use of self-explanation using a science text and the trainee identifies the strategies being used by the agents. A science text is presented on the computer screen one sentence at a time. Genie (representing a student or learner) reads the sentence aloud and produces a self-explanation. Merlin (the teacher character) asks the trainee to indicate which strategies Genie employed in producing his self-explanation. The trainee answers by clicking on a strategy in a dialog box. Merlin might then ask the student to identify and locate the various reading strategies contained in Genie's self-explanation by clicking on sentences within Genie's self-explanation. Finally, Merlin gives Genie feedback on the quality of his self-explanation. This feedback mimics the interchanges that the student will encounter in the practice module which follows the demonstration module.

In the third phase, practice, Merlin coaches and provides feedback to the trainee while the trainee practices self-explanation using the repertoire of reading strategies. The goal is to help the trainee acquire the skills necessary to integrate prior text and prior knowledge with the current sentence content. For each sentence, Merlin reads the sentence and asks the trainee to self-explain it by typing a self-explanation. The trainee types the self-explanation, and the self-explanation is evaluated. Merlin gives feedback, sometimes asking the trainee to modify unsatisfactory self-explanations. Once the self-explanation is deemed satisfactory, Merlin asks the trainee to identify what strategy was used, and Merlin provides feedback.

During this phase, the agents' interactions with the trainee are moderated by the quality of the explanation. The computational challenge is for the system to provide the student with appropriate feedback on the quality of the self-explanations within seconds. The iSTART development team has approached this evaluation challenge in four steps. First, the response is screened for metacognitive expressions. (Metacognitive expressions refer to the student's mental processes rather than to the text. E.g., "I don't understand what this text is saying.") Second, the remainder of the explanation is analyzed using both word-based and Latent Semantic Analysis (LSA) based methods [4]. Third, the results from both methods' analyses are integrated with the metacognitive screening to produce feedback in one of the following six categories:

1) response to the metacognitive content;
2) the explanation appears irrelevant to the text;
3) the explanation is too short compared to the content of the sentence;
4) the explanation is too similar to the original;
5) a hint for future self-explanations; or
6) an appropriate level of praise.

### A. iSTART Reading Strategies

In this section, the reading strategies iSTART promotes are explained. The strategies help students better understand what they read, and improve their ability to self-explain a sentence or text. These strategies are called metacognitive because they are used self consciously when self-explaining the text.

*1) Comprehension Monitoring*

Comprehension Monitoring is being aware of how well one understands what one is reading. Continuously being aware of whether one is understanding content, and if not, in what way one is having problems is the foundation of active reading.

*2) Paraphrasing*

The paraphrasing strategy requires readers to restate the sentence content in their own words. This process helps readers closely monitor their comprehension of the sentence. Paraphrasing helps readers remember the information better because the information is associated with words and phrases more familiar to them.

*3) Prediction*

Prediction is predicting what will come next in the text. Skilled readers engage in active reading in a sense that they constantly try to figure out in what direction the story or discussion in the text is developing. In addition, trying to predict the upcoming text content facilitates more close comprehension monitoring because readers compare the actual text content with the prediction.

*4) Elaboration*

Elaboration is linking information in the sentence to information you already know. Texts are almost never complete descriptions of the concepts, events or scenes they describe. Thus, comprehending text content requires a certain degree of elaboration based on an individual's knowledge. Elaboration helps relate the text content with what one already knows, thus making the text content fully integrated as part of one's existing knowledge structure.

*5) Bridging*

Bridging is linking different parts of a text together. Accurate understanding of the overall text meaning requires readers to constantly link multiple sentences in a coherent way. Identifying and understanding sentence(s) in a previous section of the text which contain the cause of the event or source of the concept described in the current sentence is important for bridging and forming a coherent understanding of overall text content.

### B. Evidence iSTART Works

Empirical studies on the effectiveness of iSTART have shown that comprehension of science texts increases in students that have been through the iSTART training. Studies at both the college ([5], [6]) and high school levels ([7], [8] [9], [10]) have indicated that iSTART improves text comprehension and strategy use over control groups. Two studies have further confirmed that iSTART training is as effective as a live, classroom-based version of the training called SERT [11], [6] from which iSTART was developed.

Research has found a pattern of results when investigating the benefits of iSTART depending on the students' prior reading skill, in that skilled readers performed better with bridging after training, whereas less skilled readers gained skills in basic text comprehension [11]. Thus, more skilled readers learned strategies that allowed them to make more connections within the text. In contrast, the less skilled readers learned the more basic level strategies (such as paraphrasing) that allowed them to make sense of the individual sentences.

The effects of practice tend to wane over time for some students [11]. These results have pointed toward a need for extended practice, and a need to improve engagement in iSTART's current practice module. MiBoard will determine whether presenting iSTART practice within a game environment will improve engagement during extended practice. MiBoard will also motivate less skilled readers to practice more effectively thus leading them to more effectively use strategies that facilitate deeper learning of textually presented material. Ultimately, MiBoard's goal is to provide a more engaging method of extended practice for all students, allowing them to further their knowledge and skills using the iSTART Reading Strategies.

## III. EDUCATION AND GAMING

MiBoard's technological goal is to build a learning environment based on serious, or educational, games. Serious games create scenarios in which the player must provide sufficient mastery of a skill normally practiced or demonstrated in an educational setting. The focus of the game should be on knowledge, not trivia or reflexes. Gredler posits five general guidelines in designing such a game [12]. First, winning the game should require the appropriate use of a specific skill and/or knowledge. Second, the content of the game should not be trivial. A serious game about biology should require knowledge about biology to win. Third, learners should not lose points for wrong answers. Penalizing wrong answers makes learners less likely to answer. Instead, the wrong answers should be identified through feedback and clarification. Fourth, games should adapt to the developmental level of the players. Games should not be too challenging or too easy. Fifth, games should not be zero-sum gains. Completion of the game should not promote a single learner as winner, but highlight the advancement that each learner obtained MiBoard currently follows all of Gredler's guidelines, except adapting to the player's developmental level to provide an engaging and pedagogical experience. Future work with MiBoard, which is outlined in the Future Work section, will include an adaptation.

Serious games should provide benefits similar to tutoring environments, such as that of iSTART. They should provide individual and adaptive learning in an environment in which learners are able to practice. Rapid feedback is essential in that it helps learners gauge their progress. Rewards and point systems are sufficient quantifiable methods of feedback beyond the traditional or verbal responses. Serious games also make practice more enjoyable for a learner. They provide a comfortable environment in which players may extrapolate existing skills or knowledge to new challenges, or even gain new perspectives on such skills.

## IV. iSTART: THE BOARD GAME

iSTART: The Board Game (iTG) was developed by Mike Rowe. iTG was created to investigate the effects of converting the practice module of iSTART to a game-based adaptation [13]. This section outlines the rules of game play in Rowe's implementation.

Rowe's game is played with 4 boards, 2 texts, 6 player tokens, 1 monster token, 120 event cards, 6 sets of 5 strategy cards, 20 task cards, and 20 power cards. One board is chosen at the beginning of a game. An event card has instructions, such as move forward 1-3 spaces, move backward 1-3 spaces, or draw a power card. Each strategy card in a set of 5 has a different strategy on it; one for each of the iSTART Strategies. A task card lists two strategies, each with an associated point value. A power card has a special power on it, such as take another turn, roll two dice, or freeze a player for 1 turn.

Each round consists of each player taking a turn reading and self-explaining, then the monster moving. During a turn, if a player is a reader, he takes a card off the Task Card deck and does not reveal it to other players. The player then reads a passage from the selected text aloud. He reads at least one sentence. For more advanced players, multiple sentences can be read. If the player is using the same text as other players, he should continue where the last reader left off, or if he is the first reader, he should select a place to begin reading. If using a different text than other readers, the player should continue where he left off, or select a place to begin reading

The reader should self-explain the text aloud, using one or both strategies on the Task Card. If the reader uses one strategy correctly, the reader gets all the points listed next to the strategy. If the reader uses both strategies correctly, the reader gets double the larger point value on the card. All other players will attempt to guess what strategy the reader used. Other players (guessers) will place one of their Strategy Cards face down in front of them, representing the strategy they think the reader used. All guessers will turn over their Strategy Cards at once. Beginning to the reader's left and continuing clockwise, each guesser should state what their guess is. If there is no disagreement, the points are awarded. If the strategy matches how the reader self-explained, and is on the Task Card, the guesser gets half the points listed next to the strategy rounded down. If the strategy matches how the reader self-explained, but is NOT on the Task Card, the guesser gets 1 point. If the strategy does not match how the reader self-explained, the guesser gets no points. If there are disagreements, do not score points until the disagreements are resolved.

To resolve disagreements, all players discuss whether the strategy use and guesses were correct, beginning with disagreements about the reader's strategy use. A majority of players must agree that the reader did not use a strategy on the task card. The reader can attempt to explain his self-explanation by showing how it was a correct use. If a majority still disagrees, the reader can try to self-explain again using the strategy for half points. If a majority of players still agrees that the reader did not use the guessed strategy, the guesser can attempt to explain why the guess is correct and where it was used in the self-explanation. If a majority still disagrees, no points are scored. After the disagreement is resolved continue clock-wise to the next disagreement and repeat the steps above.

After the discussion, the reader may now use any Power Card he has. After using a power card, or choosing not to use a power card, the reader will roll the die and move his token the number of spaces indicated on the die. The reader will then take an Event Card and perform the action on the event card. After all players have completed one turn the round ends. Roll 1 die for the monster's movement. The monster is moved half the number shown on the die rounded down.

Rowe indicated iTG was an effective form of extended practice but was not meant as a replacement for any of iSTART's existing modules. He also indicated game players found the game an enjoyable method of practicing with iSTART. Rowe theorized that a digital game would provide users another way to practice with their peers without being in the same physical location. The dissertation also mentions that the target audience of iSTART is composed primarily of students familiar with video games and that this might cause even higher levels of engagement in a computerized version than observed through the physical board game (iTG). An interactive environment also allows the possibility of adjusting the challenge of the game to the player [13].

Rowe's work concluded that this game was an effective tool for alleviating the monotony and tedium the current method of extended practice imposes upon participants

## V. MiBoard

MiBoard (Multiplayer, Interactive Board-game) is the video game version of Rowe's physical board game mentioned in the above section. MiBoard was reduced from 6 to 3 or 4 players, and the Monster was eliminated. (Rowe used the Monster as a timing mechanism. Computers provide other timing mechanisms that can be used instead of a Monster token, allowing for the game to be simplified with its removal.) In MiBoard, a task card only includes one strategy to ensure a student does not always pick the easier of the strategies with which to self-explain. All participants in MiBoard use the same text, which differs from iTG. Currently, only one board is played on in MiBoard. Future versions of MiBoard will include a wider variety of boards on which to play, as discussed in *Future Work*. Discussions and awarding points is also slightly different in MiBoard, both methods of which are described in this section.

The text used during a MiBoard game is randomly chosen from a database of science texts and includes a list of target sentences. Target sentences are the sentence numbers of the sentences to be self-explained. The text is revealed to the players gradually over the course of the game. During each turn, all sentences up to and including the next target sentence (as well as all previous sentences revealed) are shown to the players. For example, if a text has a set of target sentences 3, 5, and 6, the users see sentences 1, 2, and 3 during the first turn. During the second turn, players see sentences 1, 2, 3, 4, and 5. Varying and avoiding repeating the texts used during the game is meant to provide variety and keep the players engaged over long-term use of the game and extended practice.

The players take turns being the Reader. The other, non-Reader players are designated Guessers for that turn. On each Reader's turn, the next target sentence is revealed and the Reader is instructed to use a certain reading strategy in a self-explanation (SE) of that sentence. That strategy is randomly chosen from a list of the iSTART Reading Strategies. A point value is also randomly chosen from a list of point values including 12, 14, 16, 18, and 20. (This random selection is the automated equivalent of drawing a task card in Rowe's game.)

Upon the Reader's completion of his turn, the Guessers are shown the Reader's SE. Each Guesser decides which single strategy is most prominent in the SE and defends this choice by constructing an argument with the help of a Cascading Menu Block. (The construction of the responses is the equivalent of a player turning over his strategy card in Rowe's game.) The Cascading Menu Block (CMB) is used to provide structure for the responses (Fig 1). The CMB also reminds the students of the meanings of the strategies and how to identify their use in a self-explanation. Students are allowed a more free form exhibition of their knowledge in the Discussion that follows. Free-form responses may have worked in the supervised game conducted in Rowe's experiment, but more structure will be needed in an unsupervised environment.

Fig. 1. Strategy Identification Screen, where the player will identify the strategy used in the reader's self-explanation and then a cascading menu block is present.

The Reader uses the same device to argue that he used the specified strategy. At this point all players' arguments are revealed to all players. If there is agreement by all players about the strategy used, points are awarded to players who chose accepted strategies. Strategies are considered accepted if the majority of players chose that strategy. If the Reader's chosen strategy is deemed accepted, he is awarded the randomly selected point value. If a Guesser's chosen strategy is accepted and the strategy is the specified strategy, he is awarded half the points. If there is an accepted strategy that does not match the specified strategy, players (including the reader) who selected that strategy are awarded 5 points.

If there is disagreement on the strategy (no all of players chose the same strategy), a discussion session is initiated in which a chat-room is used by the players to express their opinions about which strategy the Reader used. Here the discussion is freed from the constraints of the CMBs but is still limited to prevent the game from turning into a social event instead of practice with the strategies. In particular, each player can make up to three contributions or he can forfeit his remaining responses. When each player has finished, forfeited his responses, or a time limit has passed, a second round of voting ensues. In this round of voting, players may select several strategies. Points are awarded to players who select accepted strategies. Again, a strategy is accepted if the majority of players select that strategy. The same scoring rules apply for this second round of voting as in the first round.

At this point the Reader rolls a die and his token is advanced along a path on the board. When the player lands on a square he draws an event card that may advance or retreat his token or allow him to draw a power card that can be used later. If the player has a power card in his possession, he may use it before rolling the die. The game is won by advancing the token to the end of the path or by accumulating enough points. The game can be reset for another, new game.

*A. Game Board*

The basic game board of MiBoard (Fig. 2.) includes the playing field, 4 player tokens, a message box, a list of players with associated scores and tokens, a button for drawing event cards, seeing the text, and getting help. The event cards cannot be drawn until after the player rolls.

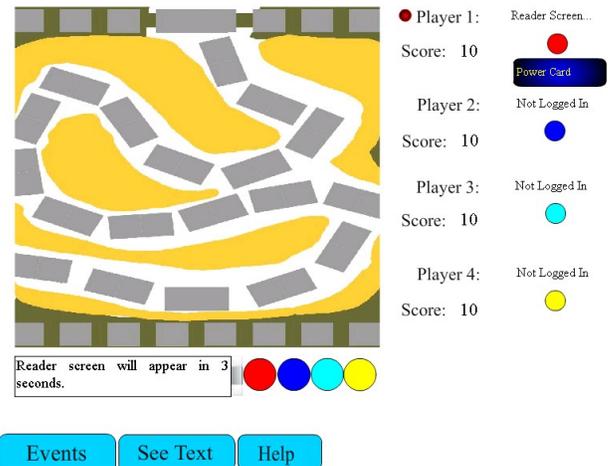

Fig. 2. Game Board show different positions/location where the player pieces can be placed. The player can see where other players are in the game.

*B. Chat*

The chat (Fig. 3) is used for the idle players to converse and for sending messages between connected players. The chat is also the medium in which players discuss disagreements in voting. The chat is only enabled during discussions and when the players are idle. In order to retain the attention of the idle players, they are allowed to chat with other idle members of the game.

*C. Reader Screen*

At the Reader Screen (Fig. 4), the Reader reads the sentence for which he is to provide a SE, and types his SE, focusing on the provided strategy. He has the option of choosing a random, new strategy or a random, new point value by clicking on the appropriately labeled buttons.

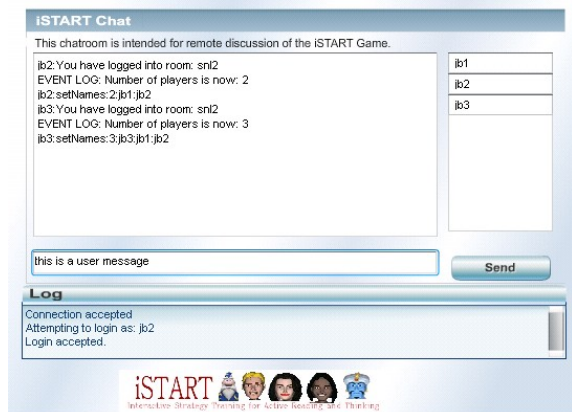

Fig. 3. iSTART Chat room allowing players to discuss their strategy selection.

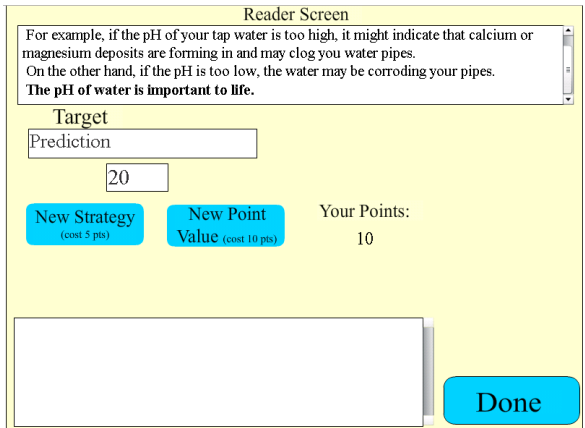

Fig. 4. Reader Screen. The reader can type in his/her self-explanation of a given text and target sentence.

### D. Guesser Screen

At the Guesser Screen (Fig. 5), players select the strategy they think was focused on by the Reader. The Guesser may only choose one such strategy at this stage in the game.

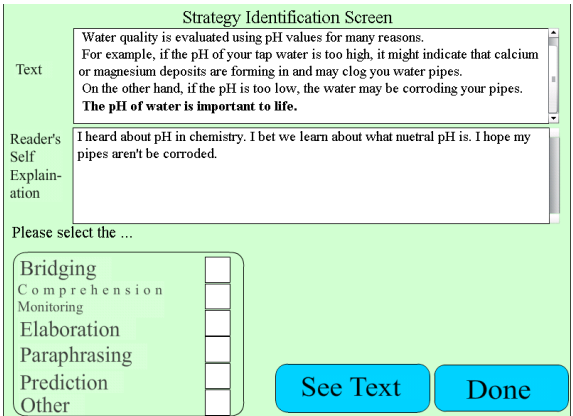

Fig. 5. Guesser Strategy Identification Screen. The remaining players will have to identify the strategy in which the reader has used in his self-explanation.

### E. Cascading Menu Block

The Cascading Menu Block (Fig. 6) is part of the Guesser Screen. It is called cascading because each time a user clicks on a check box, a new screen appears. A use is asked to click a strategy, then a reason for that selection (such as, Linked to a specific sentence), and then is asked to highlight the part of the SE in which that particular strategy was used.

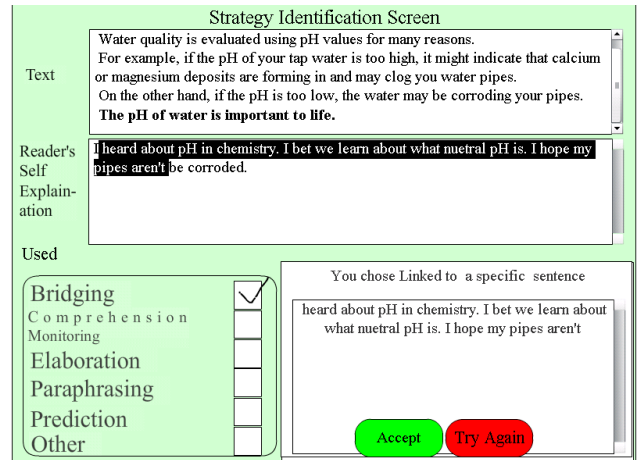

Fig. 6. Guesser Cascading Menu Block. The player not only has to identify the strategy used, but also provide the reason why he thinks the strategy was used.

### F. Summary Screen

The Summary Screen (Fig. 7) provides a summary of the explanations built by the Cascading Menu Block, as well as a summary of points earned in the round.

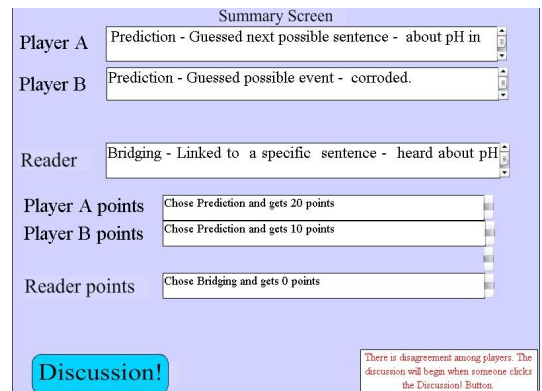

Fig. 7. Summary Screen shows selections made by all users.

### G. Discussion

The Discussion (Fig. 8) includes a set of rules (in red) and enabling of the chat room. This player has forfeited his responses by clicking the "Pass" button (which has disappeared). After the discussion, the players see the Guesser screen, where they may select as many strategies as they like. Then, the summary screen shows the new point values.

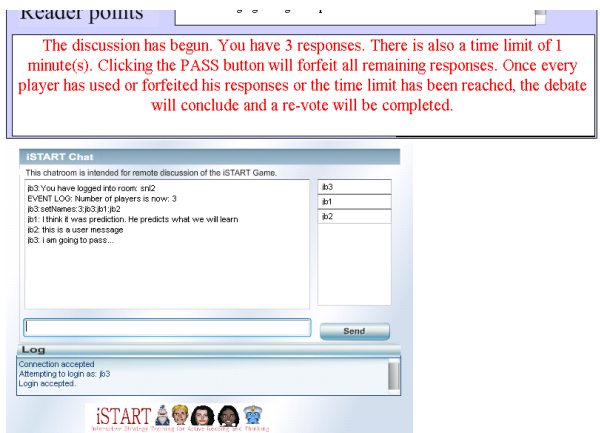

Fig. 8. Discussion through iSTART Chat. All players can discuss or argue on their strategy selections.

*H. Power Cards*

A user may use a power card by clicking on the blue power card button to bring up the power card screen (Fig. 9).

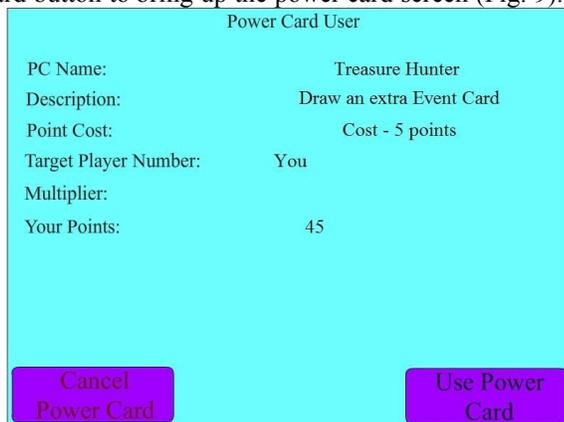

Fig. 9. Power Card available to the Reader.

VI. TECHNICAL ASPECTS AND INNOVATIONS

MiBoard was created using the Flash programming language ActionScript 3.0, JavaScript, Java Server Pages (JSP), MySQL, and ElectroServer.

ElectroServer is a multiplayer server product that facilitates interaction between many connected users and can be used for real-time audio and video streaming and recording. It is particularly useful for hosting Flash games. With its multiplayer feature, ElectroServer is suitable to be used to handle the multiplayer game in iSTART. ElectroServer works by allowing client applications, such as Flash, Java, or Silverlight, to connect to it via socket and log in. This connection is persisted as long as the client wants to stay connected. While connected, the server can push data to the client or the client can make requests of the server at any time. ElectroServer specializes in allowing communication between Flash movies through its own set of abstract data types (ADTs) and code structure.

Rooms and zones are ADTs in ElectroServer that proved particularly useful in the development of MiBoard. A room is a collection of users playing a single game that can all "see" each other. These users can easily communicate to achieve chatting or multiplayer game play. A zone is a collection of rooms. Chatting can occur as public messages sent to an entire room of users, or private messages that are sent to one or more specific users in any room.

In transforming Dr. Rowe's game, we use one room to represent a board of three or four players. The room enables each player sees to be synchronized and supports chatting among the players. Since MiBoard only supports 3 to 4 players, maximum capacities are set on each room. When a player tries to play MiBoard, he is put into a zone with a collection of rooms. MiBoard automatically searches for the first room that has not yet started and is not yet full. If there are no rooms satisfying these criteria, a new room is created and the user enters that room. Once a room has 3 players, the game may, but is not required to, begin. Beginning a game effectively prevents any further players from entering that room.

MiBoard utilizes a round-robin master-slave relationship among participating clients. Each client contains the code to run the game in its entirety. When the client is a Reader, the client controls each of the other connected clients by passing messages to each client. The clients receiving the messages and parse them to determine the desired action. These messages are sent from the chat to each client connected to the game. Upon completion of the Reader's turn, the control is passed to the next player, making his client the master, and reverting the previous master to a slave.

String parsing and recognition of user vs. game communication is essential. The messages passed have a very strict format, and cause the game to behave properly; the messages synchronize the game at each computer at which a user is playing the game. Codes are inserted into the messages for game play, and messages without codes are messages sent by the users.

The underlying infrastructure of MiBoard is particularly interesting. ActionScript 3.0 is not made to communicate with databases or other exterior entities. ActionScript only references its calling entity via the object ExternalInterface. ExternalInterface has a property "call" which tells the calling entity to invoke its function specified in the call. For example, ExternalInterface.call( "myFunc", "myParam" ) invokes the calling entity's myFunc function with the parameter myParam. Since MiBoard consists of a chat movie and a game movie imbedded in a JSP page containing JavaScript, MiBoard is able to call functions in JavaScript through the ExternalInterface object. The two movies are separate entities imbedded in the same JSP page. Therefore, one movie interacts with the other by calling functions in JavaScript that call functions in the other movie. When the board movie wants to tell all connected players that a player has moved 2 spaces forward, the board movie tells the JavaScript to tell the chat that the player just moved 2 spaces. The chat broadcasts that message to all connected players. The receiving chat movie tells its JavaScript to tell the board movie the passed message.

Logging the progress of players is essential in analyzing the effectiveness of MiBoard. This logging is done in a MySQL database. Since communication with the MySQL database occurs in iSTART's JSP pages, MiBoard must communicate within the JSP pages. JSP is a server-side language, and therefore cannot interface with the database after the page has been rendered and loaded. The method used to circumnavigate that obstacle involves the aforementioned strategy. ActionScript tells the JavaScript that it would like to log data (which is passed as a parameter to the JavaScript function). The JavaScript parses the data, and creates an iFrame of length and width 0. This invisible iFrame contains a new JSP page, which takes a MySQL query as a parameter. This new JSP page executes the passed query, and closes. This method allows ActionScript to interface with a database.

The use of ActionScript, JSP, JavaScript, and MySQL is an innovative way of linking multiple languages together, utilizing each of their unique strengths to accomplish a single, seamless system.

## VII. Future Work

MiBoard will continue to be developed and improved over the next year. The next version of MiBoard will include automatic analysis of the user self-explanations via LSA. This will ensure the self-explanations provided are accurate, and will reduce the ability of users to cheat during play. There will also be better feedback for the students after using LSA. This feedback would not replace player created feedback, but be used as a judging and correction mechanism for guesses, as well as help direct future use of the strategies in self-explanations. With the inclusion of expert feedback, the ability level of the players is no longer a limiting factor of practice.

MiBoard will also provide the ability for users to change the look and feel of the game interface which will further the appeal of the game. Different board layouts and themed skins (yet to be determined) will also be available. MiBoard will eventually be able to assess the skill level of the user and adjust the difficulty accordingly. The iSTART training modules currently implements technologies to achieve this, and extrapolating that technology into MiBoard will be beneficial to participants in the game.

## Acknowledgment

Special thanks to the iSTART Game Team at the University of Memphis under Dr. Danielle S. McNamara, and Dr. Mike Rowe for his invention.

## References

[1] D.S. McNamara, I.B. Levinstein, and C. Boonthum, "iSTART: Interactive strategy trainer for active reading and thinking," in Behavior Research Methods, Instruments, and Computers, vol. 36, 2004, pp. 222-233.
[2] M.T.H. Chi, M. Bassok, M. Lewis, P. Reimann, and R. Glaser, "Self-explanation: How students study and use examples in learning to solve problems," Cognitive Science, vol. 13, 1989, pp. 145-182.
[3] M.T.H. Chi, N. De Leeuw, M. Chiu, and C. LaVancher, "Eliciting self explanations improves understanding," Cognitive Science, vol. 18, 1994, pp.439-477.
[4] D.S. McNamara, C. Boonthum,, I.B. Levinstein, and K.K. Millis, "Evaluating self-explanation in iSTART: Comparing word-based LSA systems," in T. Landauer, D.S. McNamara, S. Dennis, and W. Kintsch eds., Handbook of Latent Semantic Analysis, Lawrence Erlbaum, Mahwah, NJ, 2007, pp. 227-241.
[5] D.S. McNamara, "SERT: Self-explanation reading training," Discourse Processes, vol. 38, 2004, pp. 1-30.
[6] T. O'Reilly, G.P. Sinclair, and D.S. McNamara, "Reading strategy training: Automated versus live," Proceedings of the 16th Annual Meeting of the Cognitive Science Society, Austin, TX: Cognitive Science Society, 2004, pp. 1059-1064.
[7] T. O'Reilly, R. Best, and D.S. McNamara, "Self-explanation reading training: Effects for low-knowledge readers," in K.Forbus, D. Gentner, and T. Regier eds., Proceedings of the 26th Annual Meeting of the Cognitive Science Society, MahWah, NJ: Erlbaum, 2004, pp. 1053-1058.
[8] T.P. O'Reilly, G.P. Sinclair, and D.S. McNamara, "iSTART: a web-based reading strategy intervention that improves students' science comprehension," in Kinshuk, D.G. Sampson, and P. Isaias eds., Proceedings of the IADIS International Conference on Cognition and Eploratory Learning in the Digital Age: CELDA, Lisbon, Portugal: IADIS Press, 2004, pp. 173-180.
[9] R.S. Taylor, T. O'Reilly, G.P. Sinclair, and D.S. McNamara, "Enhancing learning of expository science texts in a remedial reading classroom via iSTART," in S. Barab, K. Hay, and D. Hickey eds., Proceedings of the 7th International Conference of the Learning Sciences, Mahwah, NJ: Erlbaum, 2006.
[10] T. O'Reilly, R.S. Taylor, and D.S. McNamara, "Classroom based reading strategy training: Self-explanation vs. reading control," in R. Sun and N. Miyake eds., Proceedings of the 28th Annual Conference of the Cognitive Science Society, Mahwah, NJ: Erlbaum, 2006, pp 1887-1892.
[11] J.P. Magliano, S. Todaro, K.K. Millis, K. Wiemer-Hastings, H.J. Kim, and D.S. McNamara, "Changes in reading strategies as a function of reading training: A comparison of live and computerized training," Journal of Educational Computing Research, vol. 32, 2005, pp. 185-208.
[12] M.E. Gredler, "Games and simulations and their relationships to learning," in D.H. Jonassen ed., Handbook of research on educational communications and technology, Mahwah, NJ, US: Lawrence Erlbaum Assoc., 2nd ed., 2004, pp. 571-582.
[13] M. Rowe, "Alternate forms of reading comprehension strategy practice and game-based practice methods," Doctoral Dissertation, Psychology Department, the University of Memphis, 2008